# Photonics-enabled wavelet-like transform via nonlinear optical frequency sweeping and stimulated Brillouin scattering-based frequency-to-time mapping

Pengcheng Zuo [a,b], Dong Ma[a,b], and Yang Chen[a,b,*]

[a] Shanghai Key Laboratory of Multidimensional Information Processing, East China Normal University, Shanghai, 200241, China
[b] Engineering Center of SHMEC for Space Information and GNSS, East China Normal University, Shanghai, 200241, China
[*] ychen@ce.ecnu.edu.cn

**ABSTRACT**
A photonics-enabled wavelet-like transform system, characterized by multi-resolution time-frequency analysis, is proposed based on a typical stimulated Brillouin scattering (SBS) pump-probe setup using an optical nonlinear frequency-sweep signal. In the pump path, a continuous-wave optical signal is injected into an SBS medium to generate an SBS gain. In the probe path, a periodic nonlinear frequency-sweep optical signal with a time-varying chirp rate is generated, which is then modulated at a Mach–Zehnder modulator (MZM) by the electrical signal under test (SUT). The optical signal from the MZM is selectively amplified by the SBS gain and converted back to the electrical domain using a low-speed photodetector, implementing the periodic SBS-based frequency-to-time mapping (FTTM). The frequency-domain information corresponding to different periods is mapped to the time domain via the FTTM in the form of low-speed electrical pulses, which is then spliced to analyze the time-frequency relationship of the SUT in real-time. The time-varying chirp rate in each sweep period makes the signals with different frequencies have different frequency resolutions in the FTTM process, which is very similar to the characteristics of the wavelet transform, so we call it wavelet-like transform. An experiment is carried out. Multi-resolution time-frequency analysis of a variety of RF signals is carried out in a 4-GHz bandwidth limited only by the equipment.

**Keywords:** Wavelet-like transform, stimulated Brillouin scattering, microwave photonics, frequency-to-time mapping, multi-resolution time-frequency analysis.

## 1. Introduction

Microwave measurements play an important role in microwave-based applications and systems, such as electronic warfare and radar systems [1]. Time-frequency analysis (TFA) can acquire the information in two dimensions and is widely utilized for the characterization of broadband non-stationary signals. TFA methods typically include short-time Fourier transform (STFT), wavelet transform (WT), etc., which are conventionally implemented in the electrical domain [2], [3]. However, the most advanced electronic devices are still difficult to meet the needs of some high-frequency and wide bandwidth applications. For example, the sampling rate of the analog-to-digital converters and the processing speed based on digital signal processing in the electrical domain is still unsatisfactory, resulting in the analysis bandwidth and frequency of the TFA system being inherently limited [4].

Due to the huge bandwidth, high frequency, and immunity to electromagnetic interference provided by modern optics [5], photonics-assisted microwave measurements have received a lot of attention during the past two decades [6]. However, most of the photonics-based approaches mainly focus on the measurement of single-tone or multi-tone continuous-wave (CW) stationary signals. A few photonics-based TFA methods for measuring broadband non-stationary signals have been proposed [7]-[13]. Among them, a few methods for implementing STFT based on dispersive devices have been proposed [9]-[11], which generally suffer from dispersion-dependent frequency resolution and limited reconfigurability. To avoid the limitations of the existing dispersion-based STFT systems, we have proposed the stimulated Brillouin scattering (SBS)-based STFT system [12] that is free from dispersion. A periodic frequency-sweep optical signal is used to slice the signal under test (SUT) and obtain the frequency in each period, with the assistance of SBS. In each sweep period, the SUT is approximated as a stationary signal. A dynamic frequency resolution of 60 MHz and an observation bandwidth of 12 GHz are achieved and only limited by the equipment used in the experiment. However, the resolution of the SBS-based STFT system remains unchanged in one measurement although the resolution can be reconstructed in a different new measurement. Taking a 1-μs sweep period as an example, if the chirp rate of the frequency-sweep signal is relatively smaller, the system has good frequency resolution but smaller measurement bandwidth. When the chirp rate is increased, although the measurement bandwidth is increased in the same sweep period, the frequency resolution is decreased. It should be noted that we do not want to increase the measurement bandwidth by extending the period of the frequency-sweep signal because this will reduce the stationarity of the signal in a sweep period.

The resolution of STFT is globally fixed, while the WT is characterized by multi-resolution time-frequency analysis, that is, it has good time resolution (poor frequency resolution) at high frequency and good frequency resolution (poor time resolution) at low frequency. An all-optical WT system based on a dispersion-based temporal pulse shaping system is proposed in [13], which realizes multi-resolution time-frequency analysis by using multiple branches and is equivalent to performing multiple STFTs with different resolutions.

In this Letter, for the first time, a photonics-enabled wavelet-like transform system is proposed. SBS-based frequency-to-time mapping (FTTM) is periodically implemented using the optical nonlinear frequency-sweep signal. The frequency-domain information corresponding to different periods through the generated waveforms via the FTTM is formed and spliced to analyze the time-frequency relationship of the SUT in real time. The different chirp rates in a single period have a very similar function to different wavelet functions in the WT, which is the key to realizing different frequency resolutions at different frequencies. In this way, within a fixed measurement bandwidth, the measurement with variable frequency resolution can be performed, realizing functions similar to the WT. Different from the WT, a higher resolution can be distributed to the frequency band of interest by reasonably designing the chirp rate in the proposed wavelet-like transform. An experiment is performed. Multi-resolution time-frequency analysis of a variety of RF signals is carried out in a 4-GHz bandwidth, and the dynamic frequency resolution can be changed by changing the chirp rate in a single sweep period or the SBS gain bandwidth [14].

## 2. Principle

WT is one of the typical TFA methods characterized by multi-resolution time-frequency analysis. The reason why multi-resolution time-frequency analysis can be achieved is due to the use of wavelet bases with different scales. As shown in Fig. 1(a), WT has time-frequency windows of different sizes at different times and frequencies and can achieve higher frequency resolution in the low-frequency region. However, it is still limited by Heisenberg's uncertainty principle, and time resolution and frequency resolution cannot be the best of both worlds. Figure 1(b) is given to illustrate the relationship between the proposed wavelet-like transform and the WT by comparing it with Fig. 1(a).

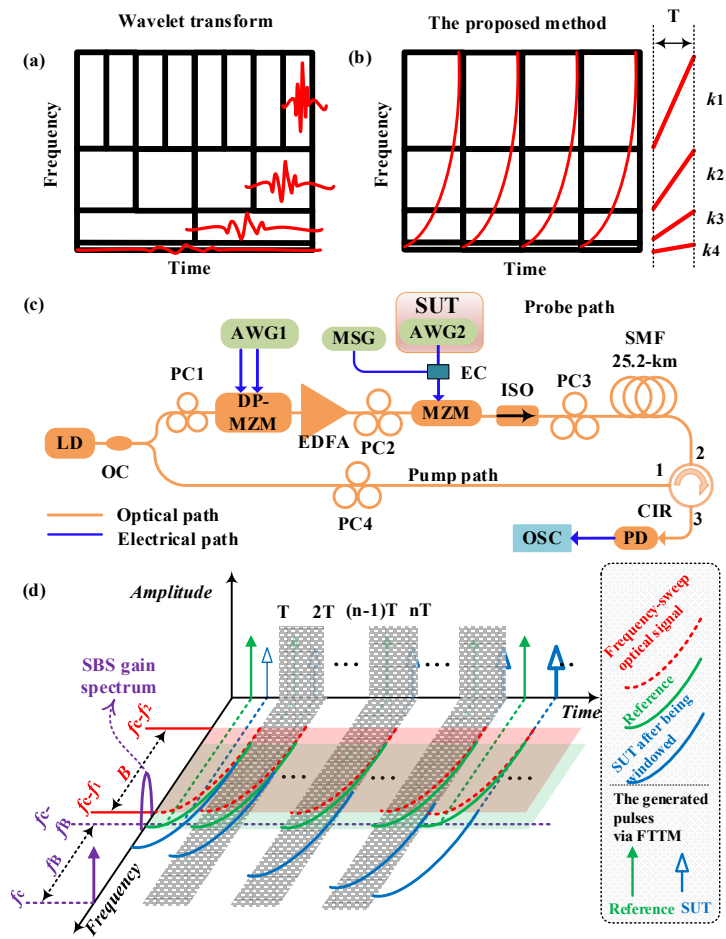

Fig. 1. Comparison of (a) the WT and (b) the proposed wavelet-like transform. (c) Schematic of the proposed wavelet-like transform system. (d) Operation principle of the wavelet-like transform. LD, laser diode; MZM, Mach–Zehnder modulator; DP-MZM, dual-parallel Mach–Zehnder modulator; AWG,

arbitrary waveform generator; MSG, microwave signal generator; EC, electrical coupler; SUT, signal under test; PC, polarization controller; EDFA, erbium-doped fiber amplifier; ISO, isolator; CIR, circulator; PD, photodetector; OSC, oscilloscope.

A proof-of-concept experiment is carried out. Figure 1(c) schematically shows the experimental setup of the wavelet-like transform system. In the lower (pump) path, a CW light wave centered at $f_c$ from a laser diode (LD, ID Photonics CoBriteDX1-1-C-H01-FA) is launched into the SBS medium, i.e., a 25.2-km single-mode fiber (SMF) in this experiment, where it interacts with the probe wave from the upper (probe) path. An SBS gain spectrum with its frequency centered at $f_c$–$f_B$ can be generated as shown in Fig. 1(d), where $f_B$ is the Brillouin frequency shift. In the probe path, a periodic nonlinear frequency-sweep optical signal is first generated via external modulation. To do so, the CW light wave is carrier-suppressed lower single-sideband modulated at a dual-parallel Mach–Zehnder modulator (DP-MZM) by a periodic nonlinear frequency-sweep electrical signal (from $f_1$ to $f_2$) from an arbitrary waveform generator (AWG1, Keysight M8195A). The time-frequency diagram is shown as the red dotted line, which has a period of $T$, a properly designed time-varying positive chirp rate of $K$, and start frequency and stop frequency of $f_c$–$f_1$ and $f_c$–$f_2$. Subsequently, the frequency-sweep optical signal is amplified by an erbium-doped fiber amplifier (EDFA, MAX-RAY EDFA-PA-35-B), and then carrier-suppressed double-sideband modulated at a Mach–Zehnder modulator (MZM, Fujitsu FTM 7938) by a combined signal. The combined signal is comprised of the SUT from AWG2 (Keysight M8190A) and a fixed single-frequency reference $f_r$ from a microwave signal generator (MSG, Agilent 83630B). In Fig. 1(d), the SUT is chosen as a linearly frequency-modulated (LFM) signal for example. Then, the output of the MZM is injected into the SMF through an optical isolator. After the SBS interaction with the pump wave, the optical signal from the circulator is detected in a photodetector (PD, Nortel PP-10G) and the generated signal is sampled by an oscilloscope (OSC, R&S RTO2032). As the sweep period $T$ is much smaller than the time duration of the SUT, the SUT is considered to be approximately stationary over a sweep period. Therefore, in each period, the frequency components of the SUT are mapped to the time domain via the SBS-based FTTM in the format of low-speed electrical pulses as shown in Fig. 1(d). Pulses in different periods represent the frequency components in the corresponding time, which are then recombined to obtain the time-frequency diagram of the SUT. The reference signal serves primarily as a label for time and frequency throughout the recombination. It should be noted that the time resolution is unchanged as the sweep period is fixed. The variation of the chirp rate in different frequencies is similar to different wavelet functions in the WT, which is the key to the multi-resolution time-frequency analysis.

In the experiment, a 25.2-km SMF is used as the SBS medium. The SBS gain spectrum is first measured by using a vector network analyzer (VNA, Agilent 8720ES). The corresponding Brillouin frequency shift $f_B$ and the 3-dB bandwidth of the SBS gain spectrum are measured to be around 10.8 GHz and 20 MHz, respectively.

## 3. Experiment and results

To preliminarily verify the multi-resolution time-frequency analysis function of the proposed wavelet-like transform system, an electrical frequency-sweep signal with a bandwidth ranging from 10.8 to 14.55 GHz and a period of 1 μs is equally divided into four segments with four different chirp rates of 1, 2, 4, and 8 GHz/μs and then applied to the system. The SUT is an LFM signal with a bandwidth ranging from 0 to 3.75 GHz and a time duration of 200 μs. Fig. 2 (a) shows the measured time-frequency diagram of the LFM signal without image registration. Because the chirp rate is changed along with time, the obtained time-frequency diagram has a nonuniform frequency axis. After registration, as shown in Fig. 2(b), the time-frequency relationship of the LFM signal is successfully recovered, which is characterized by multi-resolution time-frequency analysis, and the dynamic resolution is around 60, 120, 240, and 480 MHz in the frequency range of the four different chirp rates, respectively.

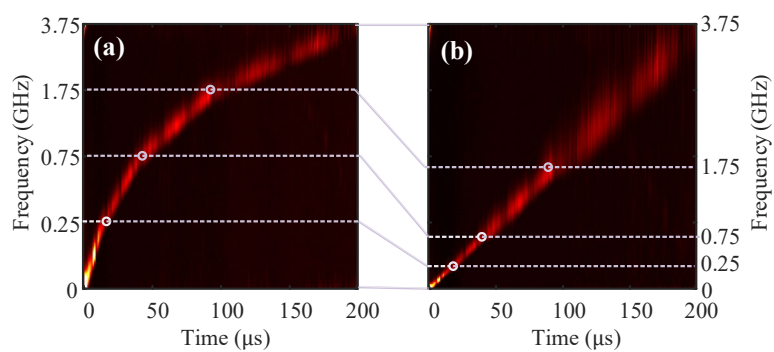

Fig. 2. Measured time-frequency diagrams of the LFM signal (a) before and (b) after image registration.

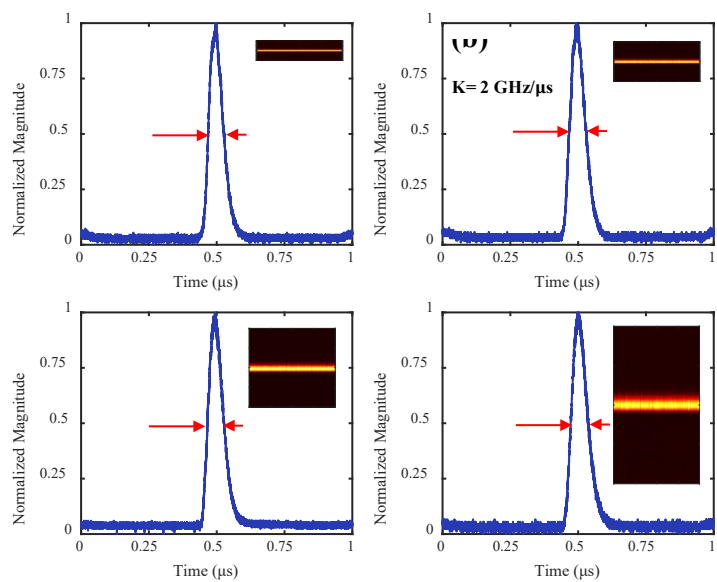

Fig. 3. Measured time-domain pulses and corresponding time-frequency diagrams of a single-tone signal by the proposed system configured with different electrical sweep signals with different chirp rates.

The reason for different resolutions under different chirp rates is further explained. A single-tone signal is measured using the proposed system under different chirp rates from 1 to 8 GHz/μs. The obtained time-domain pulses are shown in Fig. 3. It is observed that the pulse widths are all around 60 ns in these four different cases. In fact, the time-frequency diagram is the result of the recombination of two-dimensional time-domain pulses in three-dimensional space. The frequency resolution is positively correlated with the frequency range corresponding to the same pulse time width. When the chirp rate is higher, it takes less time to sweep a certain bandwidth. Therefore, the same pulse width corresponds to a larger frequency range when the chirp rate is large, that is, a worse frequency resolution.

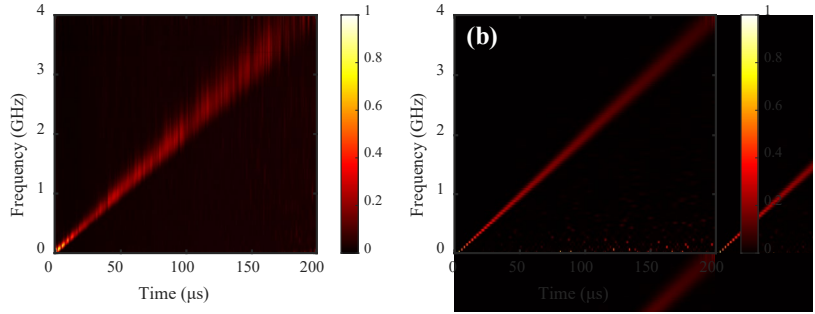

Fig. 4. (a) Measured time-frequency diagram of the LFM signal by the proposed wavelet-like transform scheme configured with an electrical frequency-sweep signal with 16 different chirp rates. (b) Simulated time-frequency diagram of the LFM signal by Matlab using WT.

It is noted that the frequency resolution of each segment shown in Fig. 2 (b) varies greatly, due to the discontinuous chirp rate change. In practical applications, the chirp rate can be continuously changed to obtain a continuously varying frequency resolution. In our experiment, For the convenience of subsequent processing and registration, another electrical frequency-sweep signal with a bandwidth ranging from 10.8 to 14.8125 GHz, a 1-μs sweep period, and 16 different chirp rates from 1 to 10 GHz/μs is used. The SUT is also a 200-μs LFM signal ranging from 0 to 4 GHz. Figure 4(a) shows the measured time-frequency diagram of the LFM signal. Compared with the result in Fig. 2(b), the obtained time-frequency diagram clearly has a more continuous frequency resolution variation in this case. For comparison, the same LFM signal is processed in Matlab using WT, with the analysis result shown in Fig. 4(b). The analysis results of the two methods have high consistency, which proves the feasibility of the proposed wavelet-like transform system.

Then, the capability of the proposed system to analyze signals in a variety of forms, including dual-chirp LFM signal (DLFM), non-linearly frequency-modulated (NLFM) signal, step-frequency (SF) signal, and frequency-hopping (FH) signal, is also studied using the same frequency-sweep signal. The time duration of the SUTs is also set to 200 μs. The signal bandwidth covers the frequency range from 0 to 4 GHz for the DLFM signal, the SF signal, and the FH signal, and from 0.5 to 4 GHz for the NLFM signal. The time-frequency analysis results for these four kinds of signals are shown in Fig. 5. As can be seen, different kinds of signals are well constructed in the time-frequency diagrams and the frequency resolution also changes with the signal frequency. The results shown in Fig. 5 verify the applicability of the wavelet-like transform proposed in this Letter in the analysis of different kinds of signals.

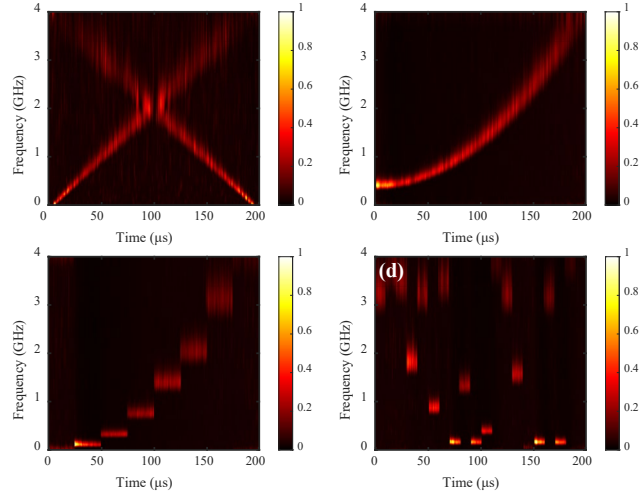

Fig. 5. Measured time-frequency diagrams by the proposed wavelet-like transform scheme for the (a) DLFM signal, (b) NLFM signal, (c) SF signal, and (d) FH signal.

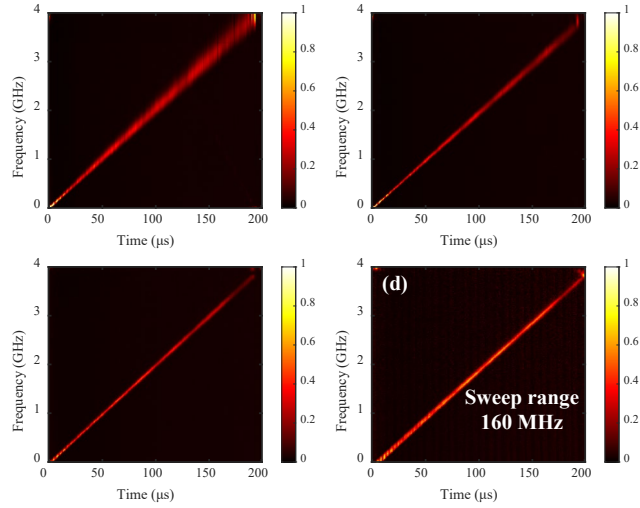

Fig. 6. Measured time-frequency diagrams of the LFM signal by the proposed wavelet-like transform scheme configured with a sweep SBS pump having a sweep bandwidth of 30, 60, 100, and 160 MHz.

In our previous work in [12] and this work, local stationary is used for the short-time Fourier transform (STFT) and wavelet-like transform. Therefore, the smaller the frequency sweep period, the better the local stationarity of the SUT. When the SBS gain bandwidth is fixed, as discussed in [12] and this Letter, the resolution decreases with the increase of the chirp rate. Although the wavelet-like transform has partly solved the contradiction between the measurement bandwidth and frequency resolution, it is highly desired that the resolution can be further improved. As discussed in [14], when the sweep signal is scanned fast enough, appropriately increasing the filter bandwidth can reduce the width of the generated pulse signal after FTTM, so better frequency measurement resolution or time-frequency analysis frequency resolution can be obtained. To broaden the SBS gain bandwidth, a DP-MZM is inserted into the pump path and modulated by frequency-sweep electrical signals centered at 6 GHz and with different sweep ranges of 30, 60, 100, and 160 MHz. To excite SBS effect with sufficient optical power, the output of the DP-MZM in the pump path is amplified by another EDFA (Amonics, EDFA-PA-35-B), with a pump power of 12.4 dBm. In this case, the Brillouin gain with increased bandwidth can be obtained [15]. The electrical frequency-sweep signal in the probe path is from 4.8 to 8.8125 GHz and has a 1-μs sweep period. An LFM signal is used as the SUT in this study. As can be seen from Fig. 6, with the increase of the sweep range of the pump wave from 30 to 100 MHz, the resolution of the results turns better, which still shows the WT characteristics that the frequency resolution is better in the low-frequency band. As the SBS gain bandwidth continues to increase to 160 MHz, the frequency resolution at higher frequencies becomes better than that at lower frequencies. The essence of this phenomenon is comprehensively discussed in [14] and will not be repeated in this Letter

## 4. Conclusion

In summary, we have proposed and experimentally demonstrated the first SBS-based wavelet-like transform system. The key to the multi-resolution time-frequency analysis is the time-varying chirp rate offered by the nonlinear frequency-sweep optical signal. The feasibility and effectiveness of the proposed system have been verified by an experiment. Multi-resolution time-frequency analysis of a variety of RF signals is carried out in a 4-GHz bandwidth. The bandwidth is determined by the sweep range of the frequency-sweep optical signal and limited only by the equipment, and the frequency resolution of the analysis is variable as the WT due to the time-varying chirp rate. The approach provides a promising alternative for microwave multi-resolution time-frequency analysis. Benefiting from the optoelectronic integration technology and chip-based SBS [16], we believe that the proposed SBS-based wavelet-like transform can better be used in real-world applications, such as electronic warfare systems. In addition, the real-time nature can be further improved by using a short high-gain SBS medium [17] or a chip-based SBS medium [16].

## Acknowledgements

This work was supported by the National Natural Science Foundation of China [grant number 61971193]; the Natural Science Foundation of Shanghai [grant number 20ZR1416100]; the Science and Technology Commission of Shanghai Municipality [grant number 18DZ2270800].

## References

1. N. Filippo, Introduction to Electronic Defense Systems, 2nd ed. (SciTech, 2006)
2. L. Cohen, "Time-frequency distributions-a review," Proc. IEEE, vol. 77, no.7, pp. 941–981, Jul. 1989.
3. V. C. Chen and H. Ling, "Joint time-frequency analysis for radar signal and image processing," IEEE Signal Process. Mag., vol. 16, no. 2, pp. 81–93, Mar. 1999.
4. B. Murmann, "ADC performance survey," 1997–2021. [Online]. Available: http://web.stanford.edu/~murmann/adcsurvey.html
5. J. Yao, "Microwave photonics," J. Lightwave Technol., vol. 27, no. 3, pp. 314–335, Feb. 2009.
6. X. Zou, B. Lu, W. Pan, L. Yan, A. Stöhr, and J. Yao, "Photonics for microwave measurements," Laser Photon. Rev., vol. 10, no. 5, pp. 711–734, Sept. 2016.

7. X. Long, W. Zou, and J. Chen, "Broadband instantaneous frequency measurement based on stimulated Brillouin scattering," Opt. Express, vol. 25, no. 3, pp. 2206–2214, Feb. 2017.

8. D. Ma, P. Zuo, and Y. Chen, "Time-Frequency Analysis of Microwave Signals Based on Stimulated Brillouin Scattering," Opt. Commun., vol. 516, Aug. 2022, Art. No. 128228.

9. M. Li and J. Yao, "All-optical short-time Fourier transform based on a temporal pulse-shaping system incorporating an array of cascaded linearly chirped fiber Bragg gratings," IEEE Photon. Technol. Lett., vol. 23, no. 20, pp. 1439–1441, Oct. 2011.

10. S. R. Konatham, R. Maram, L. R. Cortés, J. H. Chang, L. Rusch, S. LaRochelle, H. G. de Chatellus, and J. Azaña, "Real-time gap-free dynamic waveform spectral analysis with nanosecond resolutions through analog signal processing," Nat. Commun., vol. 11, Dec. 2020, Art. no. 3309.

11. X. Xie, J. Li, F. Yin, K. Xu, and Y. Dai, "STFT based on bandwidth-scaled microwave photonics," J. Lightw. Technol., vol. 39, no. 6, pp. 1680–1687, Mar. 2021.

12. P. Zuo, D. Ma, and Y. Chen, "STFT Based on stimulated Brillouin scattering," J. Lightw. Technol., vol. 40, no. 15, pp. 5052–5061, Aug. 2022.

13. M. Li and J. Yao, "Ultrafast all-optical wavelet transform based on temporal pulse shaping incorporating a 2-D array of cascaded linearly chirped fiber bragg gratings," IEEE Photon. Technol. Lett., vol. 24, no. 15, pp. 1319–1321, Jun. 2012.

14. P. Zuo, D. Ma, X. Li, and Y. Chen," Breaking the accuracy and resolution limitation of filter- and frequency-to-time mapping-based time and frequency acquisition methods by broadening the filter bandwidth," arXiv preprint arXiv: 2208.04871.

15. L. Yi, W. Wei, Y. Jaouën, M. Shi, B. Han, M. Morvan, and W. Hu, "Polarization-Independent Rectangular Microwave Photonic Filter Based on Stimulated Brillouin Scattering," J. Lightw. Technol., vol. 34, no. 2, pp. 669–675, Jan. 2016.

16. B. J. Eggleton, C. G. Poulton, P. T. Rakich, M. J. Steel, and G. Bahl, "Brillouin integrated photonics," Nat. Photon., vol. 13, pp, 664–677, Oct. 2019.

17. Kazi S. Abedin, "Observation of strong stimulated Brillouin scattering in single-mode $As_2Se_3$ chalcogenide fiber," Opt. Express, vol. 13, no. 25, pp, 10266–10271, Dec. 2005.